\documentstyle[12pt]{article}
\input{psfig}

\def\GeV{\,{\rm GeV}}

\def\sec{\,{\rm sec}}
\def\Gyr{\,{\rm Gyr}}

\def\Mpc{\,{\rm Mpc}}

\def\eV{{\,\rm eV}}

\def\cmm2{{\,\rm cm^{-2}}}
\def\cm2{{\,{\rm cm}^2}}
\def\cmm3{{\,{\rm cm}^{-3}}}
\def\gcmm3{{\,{\rm g\,cm^{-3}}}}
\def\kms{\,{\rm km\,s^{-1}}}

\def\la{\mathrel{\mathpalette\fun <}}

\def\fun#1#2{\lower3.6pt\vbox{\baselineskip0pt\lineskip.9pt
  \ialign{$\mathsurround=0pt#1\hfil##\hfil$\crcr#2\crcr\sim\crcr}}}

\begin{document}
\pagestyle{empty}
\begin{center}
\bigskip

%\rightline{FERMILAB--Pub--95/***-A}
%\rightline{astro-ph/9704024}
%\rightline{submitted to {\it }}

\vspace{.2in}
{\Large \bf COSMOLOGY 1996  }
\bigskip

\vspace{.2in}
Michael S. Turner\\

\vspace{.2in}
{\it Departments of Physics and of Astronomy \& Astrophysics\\
Enrico Fermi Institute, The University of Chicago, Chicago, IL~~60637-1433}\\

\vspace{0.1in}
{\it NASA/Fermilab Astrophysics Center\\
Fermi National Accelerator Laboratory, Batavia, IL~~60510-0500}\\

\end{center}

\vspace{.3in}
%\centerline{\bf ABSTRACT}
%\bigskip
\begin{quote}
The current state of cosmology is easy to summarize:
a very successful standard model -- the hot big-bang cosmology --
that accounts for the evolution of the Universe from
$10^{-2}\sec$ until the present; bold ideas based
upon early-Universe physics -- foremost among them
inflation and cold dark matter -- that can extend the standard
cosmology to times as early as $10^{-32}\sec$ and address
the most pressing questions; and a flood of observations --
from determinations of the Hubble constant to measurements of CBR
anisotropy -- that are testing inflation and cold dark matter.
\end{quote}

\section{Introduction}

The value of the Hubble constant has changed by about a factor of ten since
Edwin Hubble's pioneering measurements.  The context in which we
view the Universe has changed just as profoundly.  Until 1964
cosmology was mostly concerned with cosmography; the spirit of
this period was perhaps best captured by Sandage, ``the quest
for two numbers ($H_0$ and $q_0$).''  The discovery of the Cosmic
Background Radiation led to the establishment of a physical foundation
for the expanding Universe -- the hot big-bang cosmology.
The 1970s saw this model become firmly established as the standard cosmology.
In the 1980s cosmologists began trying to extend the standard cosmology
by rooting it in fundamental physics.  Inflation is a potential first
step in this program.  Today, a host of cosmological observations
are testing inflation and its cold dark matter theory of structure formation.
Although there is not agreement on how inflation is faring,
most would agree that inflation will soon be tested decisively.

\section{Foundations}

\begin{figure}[t]
\centerline{\psfig{figure=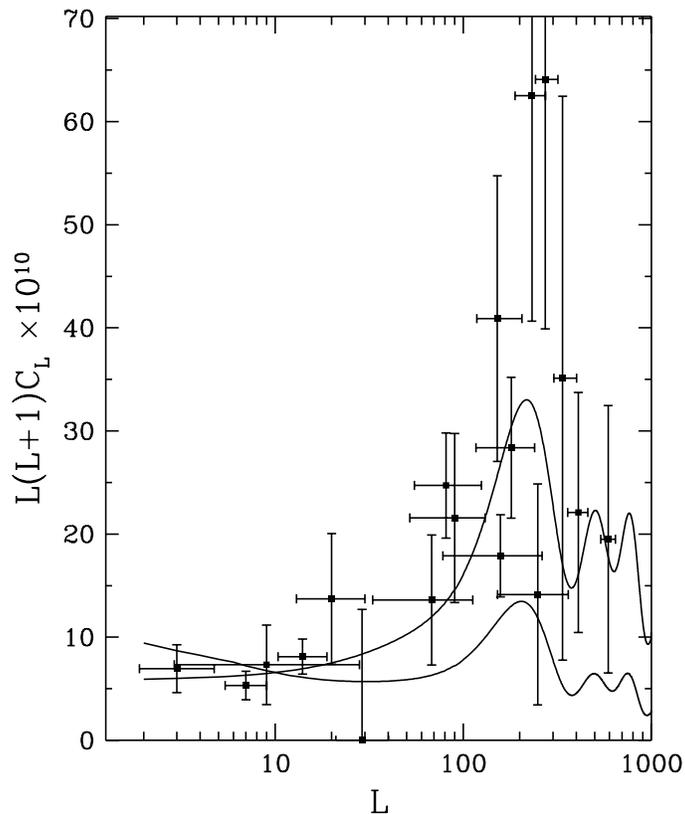,width=3.5in}}
\caption{Summary of CBR anisotropy measurements.
Plotted are the squares of the measured multipole amplitudes ($C_l = \langle
|a_{lm}|^2\rangle$) versus multipole number $l$.
The relative temperature difference on angular scale $\theta$ is
given roughly by $\protect\sqrt{l(l+1)C_l/2\pi}$
with $l\sim 200^\circ /\theta$. The theoretical curves are standard
CDM (upper curve) and CDM with $n=0.7$ (from Ref.~\protect\cite{white}).}
\end{figure}

The hot big-bang cosmology is a remarkable achievement.  It provides a
reliable accounting of the Universe from around $10^{-2}\sec$ until
the present, some $10\Gyr$ to $15\Gyr$ later.  It, together with the
standard model of particle physics and speculations about the unification
of the fundamental forces and particles, provides a firm foundation for
the sensible discussion of earlier times.

The standard cosmology rests on four observational pillars:
\begin{itemize}

\item The expansion of the Universe.  The redshifts and distances
of thousands of galaxies have been measured and are in accord with
Hubble's Law, $z = H_0 d$, a prediction of big-bang models for $z\ll 1$.

\item The Cosmic Background Radiation (CBR).  The
CBR is the most precise black body known -- deviations from
the Planck law are smaller than 0.03\% of the maximum intensity.  Its
temperature has been measured to four significant figures:
$T_0 = 2.728\pm 0.002\,$K \cite{firas}.  The only plausible origin
is the hot, dense plasma that existed in the Universe at times earlier
than $10^{13}\sec$ (epoch of last scattering and recombination).

\item Temperature fluctuations in the CBR.  Temperature differences
of order $30\mu$K between directions on the sky separated by angles
from less than one degree to ninety degrees have been measured by
more than ten different experiments \cite{white} (Fig.~1).  They establish the existence of
density inhomogeneities at the same level, $\delta \rho /\rho \sim
\delta T/T \sim 10^{-5}$, on length scales $\lambda \sim 100h^{-1}\Mpc
\,(\theta /{\rm deg})\sim 30h^{-1}\Mpc - 10^4h^{-1}\Mpc$.
Density perturbations of this amplitude, when amplified by the attractive
action of gravity over the age of the Universe, are sufficient to
explain the structure seen today.

\item Primeval abundance pattern of D, $^3$He, $^4$He and $^7$Li.
These light nuclei were produced a few seconds after the bang;
the predicted abundance pattern is consistent that seen in
primitive samples of the cosmos -- provided that the present
baryon density is between $1.5\times 10^{-31}\gcmm3$ and $4.5\times
10^{-31}\gcmm3$.  This corresponds to a fraction of critical
density $\Omega_Bh^2 = 0.008 - 0.024$ \cite{cst} (Fig.~2).
Nucleosynthesis is the earliest test of the hot big bang
and provides the best determination of the density of ordinary matter.

\end{itemize}

\begin{figure}[t]
\centerline{\psfig{figure=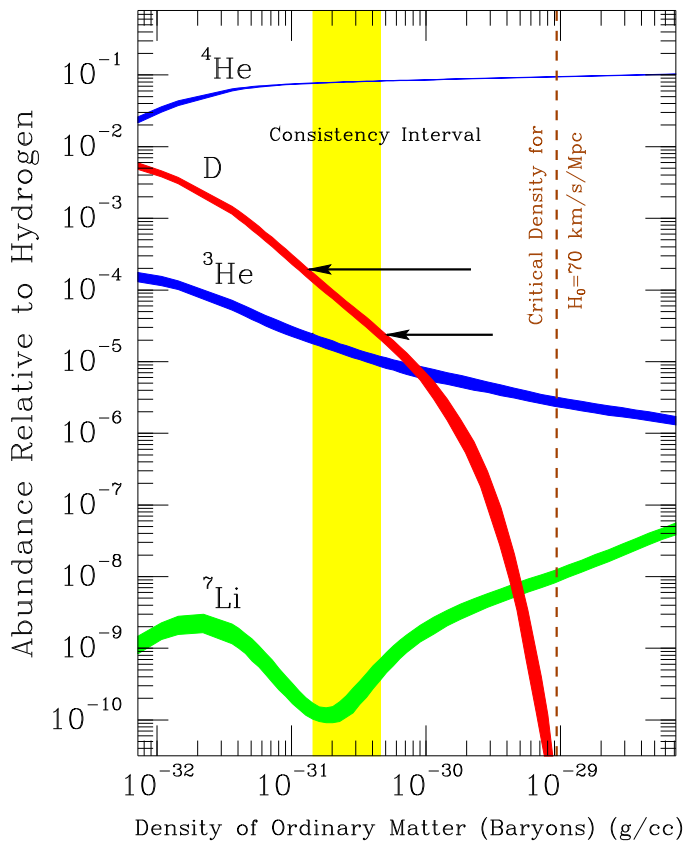,width=3.5in}}
\caption{Big-bang production of the light elements;
widths of the curves show the two-sigma theoretical uncertainty.
The primeval abundances of D, $^3$He, $^4$He and $^7$Li can be
explained if the baryon density is between $1.5\times
10^{-31}\gcmm3$ and $4.5\times 10^{-31}\gcmm3$ ($\Omega_Bh^2 =
0.008 - 0.024$).}
\end{figure}

The standard cosmology is successful in spite of our ignorance of the basic
geometry of the Universe -- age, size, and curvature -- which hinge
upon accurate measurements
of the Hubble constant and energy content of the Universe (fraction of
critical density in matter, radiation, vacuum energy, and so on).  The
expansion age, which is related to $H_0^{-1}$ and the energy content
of the Universe, is an important consistency check -- it should be
larger than the age of any object in the Universe.  The curvature radius
of the Universe is related to $H_0$ and $\Omega_0$:
$R_{\rm curv} = H_0^{-1}/\sqrt{|\Omega_0 -1|}$.

Note, the deceleration parameter is related to energy content of the
Universe, $q_0 = {1\over2} (\Omega_0 + 3\sum_i w_i\Omega_i )$, where
$\Omega_0$ is the total energy density divided the critical energy
density, $\Omega_i$ is the fraction of critical density in component
$i$ and $w_i$ is the ratio of the pressure contributed by component
$i$ to its energy density.  For a universe filled with nonrelativistic
matter, $q_0 = {1\over 2}\Omega_0$; for a universe with nonrelativistic
matter + vacuum energy (cosmological constant, $w_\Lambda = -1$),
$q_0 = {1\over 2}\Omega_0 -{3\over 2}\Omega_\Lambda$.

\section{Aspirations}

The hot big-bang model provides a firm physical basis for
the expanding Universe, but it leaves important questions unanswered.

\begin{itemize}

\item Quantity and composition of dark matter.  Most of the matter in
the Universe is dark and of unknown composition \cite{trimble}.
The peculiar velocities
of the Milky Way and other galaxies indicate that
$\Omega_{\rm Matter}$ is at least 0.3, perhaps as large as unity
\cite{pecvel}.  Luminous matter accounts for less mass density
that the lower limit to the baryon density from
nucleosynthesis ($\Omega_{\rm Lum}
\simeq 0.003h^{-1} < 0.008h^{-2} < \Omega_B$), and the upper limit to the
baryon density from nucleosynthesis is less
than 0.3 ($\Omega_B < 0.024h^{-2}<0.3$).  This defines the two dark-matter
problems central to cosmology (Fig.~3).  What is the nature of the dark
baryons? What is the nature of the nonbaryonic dark matter?

\item Formation of large-scale structure.  Gravitational
amplification of small primeval density inhomogeneities
provides the basic framework for understanding structure formation,
but important questions remain.  What is the origin of these
perturbations?  What are the details of structure formation?  The latter
is clearly tied to the dark-matter question.

\item Origin of matter-antimatter asymmetry.  During the earliest
moments ($t\la 10^{-6}\sec$), when temperatures
exceeded the rest-mass energy of nucleons, matter and antimatter existed
in almost equal amounts (thermal pair production made nucleons and
antinucleons as abundant as photons); today there is no antimatter and
relatively little matter
(one atom for every billion photons).  For this to be so, there must
have been a slight excess of matter over antimatter during the earliest
moments:  about one extra nucleon per billion nucleons and antinucleons,
for a net baryon number per photon of about $10^{-9}$.
What is the origin of this small baryon number?

\item Origin of smoothness and flatness.
Why in the large is the Universe so smooth (as evidenced by the CBR)?
The generic cosmological solutions to Einstein's equations are not smooth;
further, microphysical processes could not have smoothed things
out because the distance a light signal can travel at early times
covers only a small fraction of the Universe we can see.
Why was the Universe so flat in the beginning?  Had it not
been exceedingly flat, it would have long ago recollapsed
or gone into free expansion, resulting in a CBR temperature of
much less than 3\,K.

\item The beginning.  What launched the expansion?
What is the origin of the entropy (i.e., CBR)?  What was the big bang?
Is there a before the big bang?  Were there other bangs?
Are there more spatial dimensions to be discovered?

\end{itemize}

\begin{figure}[t]
\centerline{\psfig{figure=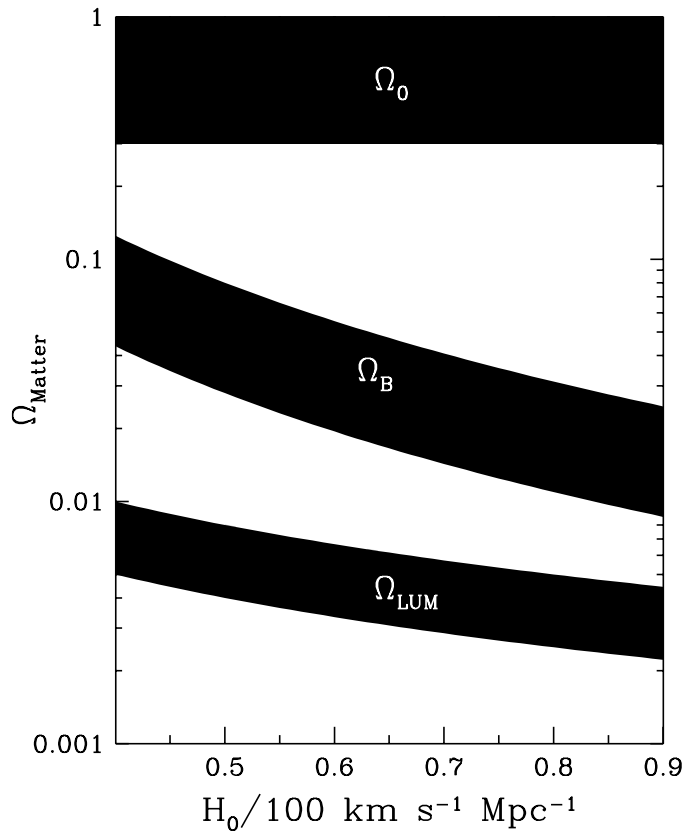,width=3.5in}}
\caption{Determinations of the matter density.
The lowest band is luminous matter, in the form of bright
stars and associated material; the middle band is the big-bang
nucleosynthesis determination of the density of baryons;
the upper band is the estimate of $\Omega_{\rm Matter}$
based upon the peculiar velocities of galaxies.  The
gaps between the bands illustrate the two dark matter problems:
most of the ordinary matter is dark and most of the matter is nonbaryonic.}
\end{figure}

This is an ambitious list.  However,
the study of the unification of the forces of Nature and the
application of these ideas to cosmology has allowed
these questions to be addressed, and many of us believe that
answers will be found in the physics of the early Universe.  Over the past
fifteen years a number of important ideas have been put forth \cite{eu}
-- baryogenesis, topological defects (cosmic strings, monopoles,
textures, and domain walls), particle dark matter, baryogenesis, and
inflation.  I will
focus on inflation -- it is the most expansive, addressing almost
all the questions mentioned above -- and is ripe for testing.

\section{Inflation and Cold Dark Matter}

Inflation \cite{inflation} holds that very early on (perhaps around $10^{-34}\sec$)
the Universe underwent a burst of exponential expansion driven by the
energy of a scalar field displaced from the minimum of its potential-energy curve.
(There are many candidates for the scalar field that drives inflation;
all involve new fields associated with physics beyond the standard
model of particle physics.)
During this growth spurt, the Universe expanded by a larger factor than
it has since.  When the scalar field evolved to the minimum of its
potential, its energy was released into a thermal bath
of particles.  This entropy is still with us today as
the Cosmic Background Radiation.

The tremendous growth in size during inflation explains the large-scale
flatness and smoothness of the Universe:
After inflation, a very tiny patch of the pre-inflationary Universe,
which would necessarily
appear flat and smooth, becomes large enough to encompass all that
we see today and more.  Since spatial curvature and $\Omega_0$ are
related, inflation predicts a critical density Universe.\footnote{Recently,
it has been shown that inflation can accommodate
$\Omega_0 <1$, but at the expense of tuning precisely
the amount of inflation \cite{loomega}.}

The most stunning prediction of inflation is the linking of large-scale structure
in the Universe to quantum fluctuations on microscopic
scales \cite{scalar} ($\ll 10^{-16}\,$cm):  The wavelengths of quantum fluctuations
in the scalar field that drives inflation are stretched to astrophysical
size by the expansion that occurs during inflation.
The continual creation of quantum fluctuations and expansion
leads to fluctuations on all length scales; they develop into density
perturbations when the vacuum energy is converted
into radiation.  The spectrum is approximately scale invariant, that is,
fluctuations in the gravitational potential that are independent
of length scale.   The overall normalization of the spectrum is
dependent upon the shape of the scalar potential, and achieving
fluctuations of the correct size to produce the observed structure
in the Universe places an important constraint on it.

An inflationary model must incorporate two other pieces of early-Universe physics:
baryogenesis \cite{baryoreviews} and particle dark matter \cite{pdm}.
Since the massive entropy released
at the end of inflation exponentially dilutes any asymmetry that might have
existed between matter and antimatter, an explanation for the
matter -- antimatter asymmetry must be provided.
Baryogenesis is an attractive one.  It holds that particle interactions
that do not conserve baryon-number and do not respect $C$ and $CP$
(matter-antimatter) symmetry occurred out-of-thermal-equilibrium and
gave rise to the small excess of matter over antimatter needed to ensure
the existence of matter today.  Details of baryogenesis remain to
be worked out and tested -- did baryogenesis occur at modest temperatures
$T\sim 200\GeV$ and involve the baryon-number violation that exists
in the standard model or did it occur at much higher temperatures
and involve grand unification physics.

Particle dark matter is necessary since inflation predicts that
the Universe is at the critical density and baryons can contribute
at most 10\% of that.  While the standard model of particle physics
does not provide a particle dark matter candidate, many theories that
attempt to unify the forces and particles predict the existence
of new, long-lived particles whose abundance today is sufficient
to provide the critical mass density.  The three most promising
candidates are:  a neutrino of mass around $30\eV$; a neutralino
of mass between $10\GeV$ and $500\GeV$ \cite{neutralino}; and an axion
of mass between $10^{-6}\eV$ and $10^{-4}\eV$ \cite{axion}.

Inflation addresses essentially all the previously mentioned questions,
including the nature of the big bang itself.  As Linde \cite{linde} has emphasized,
if inflation occurred, it has occurred time and
time again (eternally to use Linde's words). What we refer to as the big bang is
simply the beginning of our inflationary bubble, one of an infinite number
that have been spawned and will continue to be spawned ad infinitum.
From the inflationary view, there is no need for a beginning.

There is no standard model of inflation, but there are a set of
robust predictions that allow inflation to be tested.

\begin{itemize}

\item Flat Universe.  Total energy density is equal to the critical density,
$\Omega_0\equiv \sum_i \Omega_i = 1$.  Among the components $i$ are baryons, slowly
moving elementary particles (cold dark matter), radiation (a very
minor component today, $\Omega_{\rm rad}\sim 10^{-4}$), and possibly
other particle relics or a cosmological constant.

\item Approximately scale-invariant spectrum of density perturbations.
More precisely, the Fourier components of the primeval density field
are drawn from a gaussian distribution with variance given by
power spectrum $P(k) \equiv \langle |\delta_k|^2 \rangle = Ak^n$ with
$n\approx 1$ ($n=1$ is exact scale invariance),
where $k=2\pi /\lambda$ is wavenumber and the
model-dependent constant $A$ sets the overall level of inhomogeneity
and is related to the form of the inflationary potential.

\item Approximately scale-invariant spectrum of gravitational waves.
Quantum fluctuations in the space-time metric give rise to
relic gravitational waves.
The overall amplitude of the spectrum depends upon the scalar potential
in a different way than the density perturbations.
These relic gravitational waves might
be detected directly by laser interferometers that
are being built (LIGO, VIRGO, and LISA) or by the
CBR anisotropies they produce \cite{lisa}.  If the spectra of both the matter
fluctuations and gravity waves can be determined, much could be learned
about the inflationary potential \cite{recon}.

\end{itemize}

The first two predictions lead to the cold dark matter theory
of structure formation.\footnote{As a historical note the more conservative
approach of neutrino (hot) dark matter was tried first and
found to be wanting \protect\cite{nohdm}:  Since
neutrinos are light and move very fast they stream out of overdense
regions and into underdense regions, smoothing out density
inhomogeneities on small scales.  Structure forms from the top down:
superclusters fragmenting into galaxies -- which is inconsistent
with observations that indicate that superclusters are just forming today
and galaxies formed long ago.}
Within the cold dark matter (CDM) theory, there are cosmological
quantities that must be specified in order to make precise
predictions \cite{dgt}.  They can be organized into two groups.
First are the cosmological parameters:  the
Hubble constant; the density of ordinary matter; the power-law index $n$ and
overall normalization constant $A$ that quantify the density perturbations;
and the level of gravitational radiation.\footnote{The level of gravitational
radiation is important because density perturbations are normalized by
CBR anisotropy and at present it is difficult to separate the contribution
of gravity waves to CBR anisotropy from that due to density
perturbations \cite{knox}.}
(A given model of inflation predicts $A$ and $n$ as well as the level
of gravitational radiation; however, there is no standard model of inflation.
Conversely, measurements of the above quantities can constrain -- and
even be used to reconstruct -- the scalar potential that
drives inflation \cite{recon}.)

The second group specifies the composition of invisible matter
in the Universe:  radiation, dark matter, and cosmological
constant.  Radiation refers to relativistic particles:  the photons in
the CBR, three massless neutrino species (assuming none of the neutrino
species has a mass), and possibly other undetected relativistic particles.
The level of radiation is crucial since it determines when the growth
of structure begins and thereby the shape of the power spectrum
of density perturbations today.  While the bulk of the dark matter
is CDM, there could be other particle relics;
for example, a neutrino species of mass $5\eV$, which
would account for about 20\% of the critical density.

The testing of cold dark matter began more than a decade ago with
a default set of parameters (``standard CDM'')
characterized by simple choices for both
the cosmological and the invisible matter parameters:
precisely scale-invariant density perturbations ($n=1$), $h=0.5$,
$\Omega_B =0.05$, $\Omega_{\rm CDM}=0.95$;
no radiation beyond photons and three massless neutrinos; no
dark matter beyond CDM; no gravitational waves; and zero cosmological constant.
The overall level of the matter inhomogeneity -- set by the
constant $A$ -- was fixed by comparing the predicted level of inhomogeneity
today with that seen in the distribution of bright galaxies.
Bright galaxies may or may not faithfully trace the distribution of mass.
In fact, there is some evidence that bright galaxies are
more clustered than mass, by a factor called the bias, $b \simeq 1 - 2$.
The distribution of galaxies today only fixes $A$ up to the bias factor $b$.

An important change occurred with the detection of CBR anisotropy
by COBE in 1992 \cite{dmr}.  The COBE measurement permitted a precise
determination of the amplitude of density perturbations on very large scales,
without regard to biasing.  (The CBR anisotropy detected by COBE
arises mainly from density fluctuations on scales of around $10^3h^{-1}\Mpc$.)
There was a surprise:  For standard CDM, the COBE normalization
predicts too much power on the scales of clusters and smaller \cite{jpo-ll}.

Figure 4 illustrates clearly that this problem simply reflects
a poor choice for the standard parameters.  It shows that there are
many COBE-normalized CDM models that are consistent with
measurements of the large-scale structure that exists today
(shape of the power spectrum of the galaxy distribution,
abundance of clusters, and early formation of structure in the
form of damped Lyman-$\alpha$ clouds; see Ref.~\cite{dgt}).  Organized
into families characterized by their invisible matter content they are:
CDM + cosmological constant ($\Lambda$CDM) \cite{lambda},
CDM + a small amount of hot dark matter ($\nu$CDM) \cite{nucdm},
CDM + additional relativistic particles ($\tau$CDM) \cite{taucdm},
and CDM with standard invisible matter content \cite{h30,stp}.

\begin{figure}[t]
\centerline{\psfig{figure=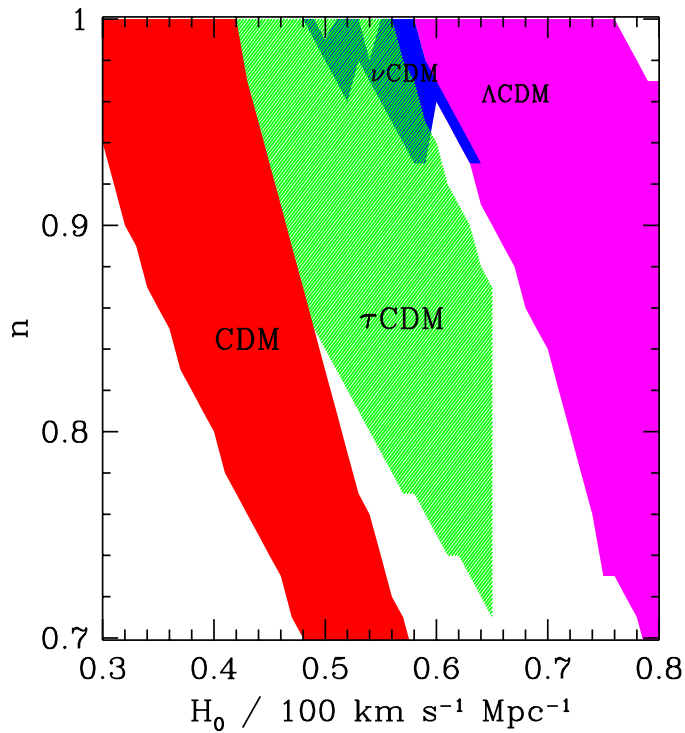,width=3.5in}}
\caption{Acceptable values of the cosmological
parameters $n$ and $h$ for CDM models with standard
invisible-matter content (CDM), with 20\% hot dark matter ($\nu$CDM),
with additional relativistic particles (the energy equivalent
of 12 massless neutrino species, denoted $\tau$CDM), and with a cosmological
constant that accounts for 60\% of the critical density ($\Lambda$CDM).
The $\tau$CDM models have been truncated at a Hubble constant
of $65\kms\Mpc^{-1}$ because a larger value would result in a
Universe that is younger than $10\Gyr$ (from Ref.~\protect\cite{dgt}).}
\end{figure}

\begin{figure}[t]
\centerline{\psfig{figure=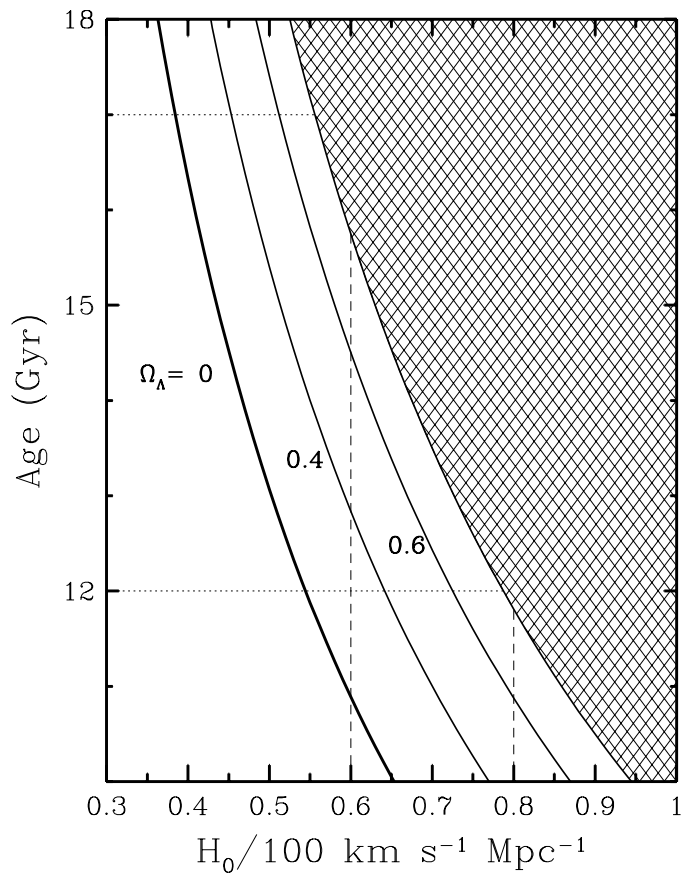,width=3.5in}}
\caption{The relationship between age and $H_0$ for flat-universe
models with $\Omega_{\rm Matter} = 1 - \Omega_\Lambda$.
The cross-hatched region is ruled out because
$\Omega_{\rm Matter} < 0.3$.  The broken lines indicate
the favored range for $H_0$ and for the age of the Universe.}
\end{figure}

There are two additional pieces of data that have
significant leverage on CDM models:  The Hubble constant/age of the Universe
and the cluster baryon fraction.  Determinations of
the Hubble constant based upon a variety of techniques (Type Ia and II
supernovae, IR Tully-Fisher and fundamental plane methods)
have converged on a value between $60\kms\Mpc^{-1}$ and $80\kms\Mpc^{-1}$ \cite{h_0}.
This corresponds to an expansion age of less than $11\Gyr$ for a flat,
matter-dominated model; for $\Lambda$CDM, the expansion age can be
significantly higher, as large as $16\Gyr$ for $\Omega_\Lambda = 0.6$ (Fig.~3).
On the other hand, the ages of the oldest globular clusters
indicate that the Universe is between $13\Gyr$ and $17\Gyr$ old;
further, these age determinations, together with the
those for the oldest white dwarfs and the long-lived radioactive elements,
provide an ironclad case for a Universe that is at least $10\Gyr$ old
\cite{age}.  Unless the age of the Universe and the
Hubble constant are near the lowest values consistent with current
measurements, only $\Lambda$CDM model is viable.

Clusters are large enough
that the baryon fraction should reflect its universal value,
$\Omega_B/\Omega_{\rm Matter} = (0.008 - 0.024)h^{-2}/(1-\Omega_\Lambda )$.
Most of the (observed) baryons in clusters are in the hot,
intracluster x-ray emitting gas.  From x-ray measurements of
the flux and temperature of the gas, baryon fractions
in the range $(0.04 - 0.10)h^{-3/2}$ have been
inferred \cite{gasratio}; further,
a recent detailed analysis and comparison to numerical models
of clusters in CDM indicates
an even smaller scatter, $(0.07\pm 0.007)h^{-3/2}$ \cite{evrard}.
From the cluster baryon fraction and $\Omega_B$,
$\Omega_{\rm Matter}$ can be inferred:  $\Omega_{\rm Matter}
= (0.25\pm 0.15)h^{-1/2}$, which, assuming that $h\ge 0.5$, implies
$\Omega_{\rm Matter} \le 0.35 \pm 0.2$.  Unless one of the assumptions
underlying this analysis is wrong, only $\Lambda$CDM is viable.

At the moment, the observations point to $\Lambda$CDM as the
best fit CDM model \cite{bestfit} (Fig.~6).  The existence of
a cosmological constant raises a fundamental
issue -- the origin of the implied vacuum energy density.
However, one should bear in mind that the case for $\Lambda$CDM
hinges upon the Hubble constant and cluster baryon fraction,
and neither measurement is completely settled.

\begin{figure}[t]
\centerline{\psfig{figure=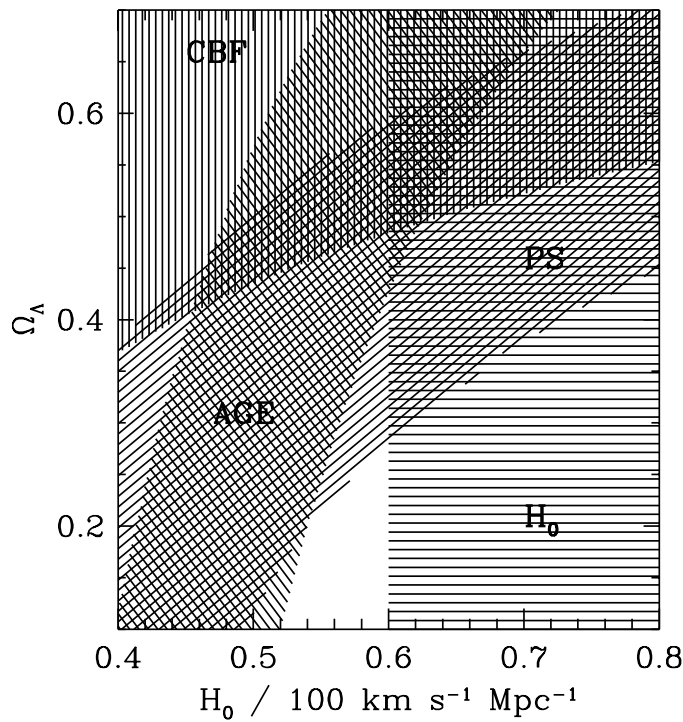,width=3.5in}}
\caption{Summary of constraints projected onto
the $H_0$ -- $\Omega_{\rm Matter}$ plane:
(CBF) comes from combining the BBN limit to the
baryon density with x-ray observations
of clusters; (PS) arises from the power spectrum;
(AGE) is based on age determinations of the Universe;
($H_0$) indicates the range currently favored for the
Hubble constant.  (Note the constraint $\Omega_\Lambda <0.7$
has been implicitly taken into account since the
$\Omega_\Lambda$ axis extends only to 0.7.)
The darkest region indicates the parameters
allowed by all constraints (from \protect\cite{bestfit1}.)}
\end{figure}

\section{Concluding Remarks}

At the moment, $\Lambda$CDM best accommodates all the observations, but
I believe the evidence is not yet strong enough to abandon the
other CDM models.  Especially because additional observations
will soon be able to decisively distinguish between the
different models as well as testing inflation.  They include:

\begin{itemize}

\item Deceleration parameter.  $\Lambda$CDM
predicts $q_0 \equiv {1\over 2} - {3\over 2}\Omega_\Lambda\sim
-{1\over 2}$, while the other CDM models predict $q_0 = {1\over 2}$.
Two groups (The Supernova Cosmology Project and The High-z Supernova Team)
are hoping to determine $q_0$ to a precision of $\pm 0.2$ by
using distant Type Ia supernovae ($z\sim 0.3 - 0.7$) as standard candles.
Together, they discovered more than 40 high redshift supernovae
last fall and winter and both groups should be announcing results soon.

\item Hubble constant.  Since the Universe is at least $10\Gyr$ old,
a determination that the Hubble constant is $65\kms\Mpc^{-1}$ or greater
would rule out all models but $\Lambda$CDM; a determination that
the Hubble constant is greater than $60\kms\Mpc^{-1}$ would
require nonminimal invisible matter content (e.g., some hot dark
matter or extra radiation); a value less than $60\kms\Mpc^{-1}$
would make the simplest CDM models viable.

\item Cluster baryon fraction.  This strongly favors
$\Lambda$CDM.  Further evidence that x-ray measurements
have correctly determined the total cluster mass (e.g., from weak
gravitational lensing) and baryon mass (e.g., from AXAF) would
strengthen the case for $\Lambda$CDM.  On the other hand,
discovery of a systematic effect
that lowers the cluster baryon fraction by a factor of two
(e.g., underestimation of cluster mass because gas is not supported
by thermal pressure alone, or overestimation of cluster gas mass
because the gas is clumped) would undermine the case for $\Lambda$CDM.

\item Evolution of Structure.
The study of the Universe at high redshift by HST and Keck
will test CDM and distinguish between models.  For example,
$\Lambda$CDM predicts earlier structure formation, while $\nu$CDM
predicts later structure formation.

\item Redshift surveys. Two large redshift surveys are coming on
line.  The Sloan Digital Sky Survey with gather a million redshifts
over a quarter of the sky; the Two-degree Field will gather 250,000
redshifts in hundreds of fields that are two degrees across.  Both
should be able to discriminate between the different CDM models by
better measuring the power spectrum of inhomogeneity and other related
quantities.

\item Determination of the primeval deuterium abundance.
A definitive measurement of the deuterium abundance in a high-redshift
hydrogen cloud can be used to determine the primeval deuterium
abundance and thereby the baryon density ($\Omega_Bh^2$)
to a precision of 10\% or so.  Such a measurement would pin down this important
cosmological parameter, sharpen the cluster baryon fraction test,
and, when the baryon density is determined from CBR anisotropy, provide
a consistency test of the standard cosmology.  Results -- though
not a consensus -- for
the deuterium abundance in high-redshift hydrogen clouds ($z\sim
2.5 -4.6$) have been reported \cite{deuterium};
further observations with the Keck and the HST should clarify matters
and lead to a 10\% determination of the primeval deuterium abundance.

\item Laboratory search for particle dark matter.  An experiment with
sufficient sensitivity to detect axions in the halo of the Milky Way
is now taking data \cite{axion2}; several experiments that can detect neutralinos
will start operating soon.  In addition, a host of experiments to search
for evidence of neutrino mass are underway (e.g., at Los Alamos, CERN,
Fermilab, Kamiokande and other laboratories).

\item CBR anisotropy in the MAP/COBRAS/SAMBA era.  Last, but certainly
not least, the high-resolution
CBR maps that will be made by these two satellite-borne experiments
as well as ground and balloon based experiments
will test both inflation and CDM decisively.  By measuring the multipole
amplitudes out to $l\sim 3000$ they will be able to simultaneously
determine $h$, $\Omega_0$, $\Omega_\Lambda$, $\Omega_Bh^2$, $\Omega_\nu$,
$n$, and $T/S$ to good precision (better than 10\%) (Fig.~7)
\cite{learncbr}.

\end{itemize}

\begin{figure}[t]
\centerline{\psfig{figure=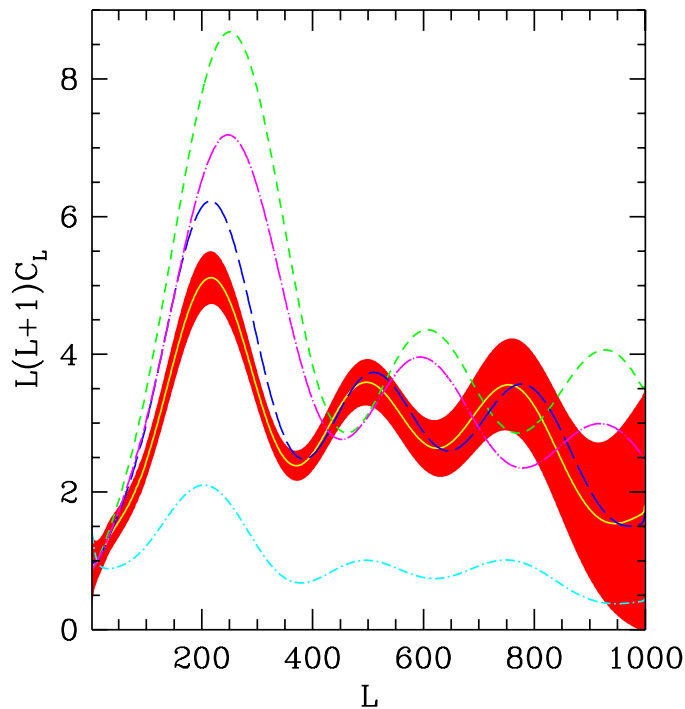,width=3.5in}}
\caption{Angular power spectra of
CBR anisotropy for several CDM models and the anticipated
uncertainty (per multipole) from a CBR satellite experiment
similar to MAP.   From top to bottom the CDM models
are:  CDM with $h=0.35$, $\tau$CDM with the energy equivalent
of 12 massless neutrino species, $\Lambda$CDM with $h=0.65$ and
$\Omega_\Lambda = 0.6$, $\nu$CDM with $\Omega_\nu = 0.2$,
and CDM with $n=0.7$ (from Ref.~\protect\cite{dgt}).}
\end{figure}

This is an exciting time in cosmology.  We have a successful
standard model, in inflation a bold and expansive paradigm for extending it,
and a flood of observations to test paradigm.
Soon we will know if inflation and
cold dark matter are to become part of the standard cosmology, and
whether or not we are living in the golden age of cosmology.

\paragraph{Acknowledgments.}  This work was supported by the DoE (at Chicago
and Fermilab) and by the NASA (at Fermilab by grant NAG 5-2788).


\begin{thebibliography} {cst}

\bibitem{firas} J. Mather et al., {\it Astrophys. J.} {\bf 420},
439 (1994); D.J.~Fixsen et al., {\it ibid}, in press (1996).

\bibitem{white} For example see, M.~White, D.~Scott, and J.~Silk,
{\it Science} {\bf 268}, 829 (1995).

\bibitem{cst} For example see, C.~Copi, D.N.~Schramm, and M.S.~Turner,
{\it Science} {\bf 267}, 192 (1995); {\it Phys. Rev. Lett.} {\bf 75}, 3981
(1995); N.~Hata et al., {\it ibid}, 3977 (1995).

\bibitem{trimble} V.~Trimble, {\it Ann. Rev. Astron.
Astrophys.} {\bf 25}, 425 (1987). 

\bibitem{pecvel} M.~Strauss and J.~Willick, {\it Phys. Repts.}
{\bf 261}, 271 (1995); A.~Dekel, {\it Ann. Rev. Astron. Astrophys.}
{\bf 32}, 319 (1994).

\bibitem{eu} For example see, E.W.~Kolb and M.S.~Turner, {\it The Early
Universe} (Addison-Wesley, Redwood City, CA, 1990).

\bibitem{inflation} A.H.~Guth, {\it Phys. Rev. D} {\bf 23}, 347 (1981);
M.S.~Turner, {\it Ann. NY Acad. Sci.} {\bf 759}, 153 (1995).

\bibitem{loomega} M.~Bucher A.S.~Goldhaber, and N.~Turok, {\it Phys. Rev. D}
{\bf 52}, 3314 (1995); M.~Bucher and N.~Turok, hep-ph/9503393.

\bibitem{scalar} A. H. Guth and S.-Y. Pi, {\it Phys. Rev. Lett.}
{\bf 49}, 1110 (1982); S. W. Hawking, {\it Phys. Lett. B} {\bf 115}, 295
(1982); A. A. Starobinskii, {\it ibid} {\bf 117}, 175 (1982);
J. M. Bardeen, P. J. Steinhardt, and M. S. Turner, {\it Phys. Rev. D}
{\bf 28}, 697 (1983).

\bibitem{baryoreviews} E.W.~Kolb and M.S.~Turner,
{\it Ann. Rev. Nucl. Part. Sci.} {\bf 33}, 645 (1983);
A.~Dolgov, {\it Phys. Repts.} {\bf 222}, 309 (1992);
A.~Cohen, D.~Kaplan, and A.~Nelson, {\it Ann. Rev. Nucl.
Part. Sci.} {\bf 43}, 27 (1992).

\bibitem{pdm} M.S.~Turner, {\it Proc. Nat. Acad. Sci. (USA)} {\bf 90},
4827 (1993).

\bibitem{neutralino} For example see, G.~Jungman, M.~Kamionkowski,
and K.~Griest, {\it Phys. Rept.} {\bf 267}, 195 (1996).

\bibitem{axion} For example see, M.S.~Turner, {\it Phys. Rept.}
{\bf 197}, 17 (1990).

\bibitem{linde} A.D.~Linde, {\it Inflation and Quantum Cosmology}
(Academic Press, San Diego, CA, 1990).

\bibitem{lisa} M.S.~Turner, astro-ph/9607066.

\bibitem{recon} M.S.~Turner, {\it Phys. Rev. D} {\bf 48}, 3502 (1993);
E.~Copeland et al., {\it Rev. Mod. Phys.}, in press (1996).

\bibitem{nohdm} S.D.M.~White, C.~Frenk and M.~Davis, {\it Astrophys. J.}
{\bf 274}, L1 (1983).

\bibitem{knox} L.~Knox and M.S.~Turner, {\it Phys. Rev. Lett.}
{\bf 73}, 3347 (1994).

\bibitem{jpo-ll} J.P.~Ostriker, {\it Ann. Rev. Astron. Astrophys.}
{\bf 31}, 689 (1993); A.~Liddle and D.~Lyth,  {\it Phys. Repts.} {\bf 231},
1 (1993).

\bibitem{dgt} For example see, S.~Dodelson, E.~Gates and M.S.~Turner,
{\it Science}, in press (astro-ph/9603081).

\bibitem{dmr} G.F.~Smoot et al., {\it Astrophys. J.} {\bf 396}, L1 (1992);
C.L.~Bennett et al., {\it Astrophys. J.} {\bf 464}, L1 (1996);
K.M.~Gorski et al., {\it Astrophys. J.} {\bf 464}, L11 (1996); M. White and
E.F.~Bunn, {\it ibid} {\bf 450}, 477 (1995).

\bibitem{lambda} M.S.~Turner, G.~Steigman, and L.~Krauss,
{\it Phys. Rev. Lett.} {\bf 52}, 2090 (1984); M.S.~Turner,
{\it Physica Scripta} {\bf T36}, 167 (1991); P.J.E.~Peebles,
{\it Astrophys. J.} {\bf 284}, 439 (1984); G.~Efstathiou et al.,
{\it Nature} {\bf 348}, 705 (1990); L.~Kofman and A.A.~Starobinskii,
{\it Sov. Astron. Lett.} {\bf 11}, 271 (1985); L.~Krauss and M.S.~Turner,
{\it Gen. Rel. Grav.} {\bf 27}, 1137 (1995); J.P.~Ostriker and
P.J.~Steinhardt, {\it Nature} {\bf 377}, 600 (1995).

\bibitem{nucdm} Q.~Shafi and F.~Stecker, {\it Phys. Rev. Lett.} {\bf 53},
1292 (1984); M.~Davis, F.~Summers, and D.~Schlegel, {\it Nature} {\bf 359},
393 (1992); J.~Primack et al., {\it Phys. Rev. Lett.} {\bf 74}, 2160 (1995);
D.~Pogosyan and A.A.~Starobinskii, {\it Astrophys. J.} {\bf 447}, 465 (1995).

\bibitem{taucdm} S.~Dodelson, G.~Gyuk, and M.S.~Turner, {\it Phys. Rev. Lett}
{\bf 72}, 3578 (1994); J.R.~Bond and G.~Efstathiou, {\it Phys. Lett. B}
{\bf 265}, 245 (1991).

\bibitem{h30} J.~Bartlett, A.~Blanchard, J.~Silk,
and M.S.~Turner, {\it Science} {\bf 267}, 980 (1995).

\bibitem{stp} R.~Cen, N.~Gnedin, L.~Kofman, and
J.P.~Ostriker, {\it ibid} {\bf 399}, L11 (1992);
R.~Davis et al., {\it Phys. Rev. Lett.}
{\bf 69}, 1856 (1992); F.~C.Adams, J.R.~Bond,  K.~Freese, J.A.~Frieman,
and A.~Olinto, {\it Phys. Rev.} {\bf D47}, 426 (1993);
M.~White, D.~Scott, J.~Silk, and M.~Davis,
{\it Mon. Not. Roy. Astron. Soc.} {\bf 276}, L69 (1995).

\bibitem{h_0} A.~Reiss, R.P.~Krishner, and W.~Press, {\it Astrophys. J.}
{\bf 438}, L17 (1995); M.~Hamuy et al, {\it Astron. J.} {\bf 109}, 1 (1995);
W.~Freedman et al., {\it Nature} {\bf 371}, 757 (1994).

\bibitem{age} B.~Chaboyer, P.~Demarque, P.J.~Kernan, and L.M.~Krauss,
{\it Science} {\bf 271}, 957; M.~Bolte and C.J.~Hogan, {\it Nature}
{\bf 376}, 399 (1995); J.~Cowan, F.~Thieleman, and J.W.~Truran, {\it
Ann. Rev. Astron. Astrophys.} {\bf 29}, 447 (1991).

\bibitem{gasratio} S.D.M.~White et al., {\it Nature} {\bf 366},
429 (1993); U.G.~Briel et al., {\it Astron. Astrophys.} {\bf 259},
L31 (1992); D.A.~White and A.C.~Fabian, {\it Mon. Not. Roy. Astron.
Soc.} {\bf 273}, 72 (1995); A.E.~Evrard, C.A.~Metzler,
and J.F.~Navarro, {\it Astrophys. J.} in press (1996).

\bibitem{evrard} G.~Evrard, in preparation (1996).

\bibitem{bestfit} L.~Krauss and M.S.~Turner, {\it Gen. Rel. Grav.} {\bf 27},
1137 (1995); J.P.~Ostriker and P.J.~Steinhardt, {\it Nature} {\bf 377}, 600 (1995).

\bibitem{bestfit1} L.~Krauss and M.S.~Turner, {\it Gen. Rel. Grav.} {\bf 27},
1137 (1995).

\bibitem{deuterium} A.~Songaila et al., {\it Nature} {\bf 368}, 599 (1994);
R.F.~Carswell et al., {\it Mon. Not. Roy. astron. Soc.} {\bf 268}, L1 (1994);
M.~Rugers and C.J.~Hogan, {\it Astrophys. J.} {\bf 459}, L1
(1996); D.~Tytler, X.-M.~Fan and S.~Burles, {\it Nature} {\bf 381},
207 (1996); D.N.~Schramm and M.S.~Turner, {\it Nature}
{\bf 381}, 193 (1996); S.~Burles and D.~Tytler, {\it Science}, in press (1996)
(astro-ph/9603070); E.J.~Wampler et al., {\it Astron. Astrophys.},
in press (1996) (astro-ph/9512084); R.F.~Carswell et al., {\it Mon.
Not. R. astron. Soc.} {\bf 278}, 506 (1996); M.~Rugers and C.J.~Hogan,
astro-ph/9603084); L.~Cowie and A.~Songaila, in preparation (1996).

\bibitem{axion2} C.~Hagmann et al., astro-ph/9607022.

\bibitem{learncbr}  For example see,
L.~Knox, {\it Phys. Rev. D} {\bf 52}, 4307 (1995);
G.~Jungman, M.~Kamionkowski, A.~Kosowsky, and D.~Spergel,
{\it Phys. Rev. D} {\bf 54}, 1332 (1996).

\end{thebibliography}
\end{document}